\documentclass[letterpaper,12pt]{article}
\usepackage{tabularx} 
\usepackage{amsmath}  
\usepackage{graphicx} 
\usepackage[margin=1in,letterpaper]{geometry} 
\usepackage{epstopdf}
\usepackage{cite} 
\usepackage[final]{hyperref} 
\hypersetup{
	colorlinks=true,       
	linkcolor=blue,        
	citecolor=blue,        
	filecolor=magenta,     
	urlcolor=blue         
}

\begin{document}

\title{{Model for Diffusion Limited Crystal Growth with and without Growth Rate Dispersion}
\author{Douglas A. Barlow \footnote{dabarlow@sewanee.edu}\footnote{doug.barlow@protonmail.com}~,~ Kylene Monaghan\\
\small Department of Physics, Woods Laboratories, The University of the South, Sewanee TN, 37383 USA\\}}

\date{}
\maketitle

\begin{abstract}
We show, in this report, how a population balance model differential equation describing batch crystal growth from solution can be solved in closed form for the case of diffusion limited growth with and without modeling the effects of growth rate dispersion. By letting the growth rate diffusivity be directly proportional to the supersaturation, a closed form solution can be found for the case of a separable distribution function of crystal sizes. This result requires that the ratio of the solute diffusion coefficient to the growth rate diffusivity coefficient, be restricted to odd integer values. This implies that when growth rate dispersion is present, varying the conditions of the solution can only lead to one or more equilibrium size distributions out of an infinite discrete set of possible size distribution functions. We deal with this restriction by letting the growth rate diffusivity coefficient be restricted to a spectrum of discrete values. Results are compared with experimental data for two cases of lactose crystal growth from solution, and one for sucrose, where it was suggested that there are two distinct types of kinetic behavior in the growing crystal ensemble; one of slow growing crystals and the other fast growing. Our model is able to describe the equilibrium distribution of these crystals as being composed of a combination of two size distribution functions, each with its own uniquely valued growth rate diffusivity coefficient.
\end{abstract}


\clearpage

\section{Introduction}

Growth rate dispersion, (GRD), where crystals of identical size and material in the same solution have different growth rates, was identified during the batch growth of sucrose from solution and reported in the literature as early as 1969 \cite{White1, White2}. Since that time, this effect has been observed in a variety of systems and several explanations have been given for the probable cause \cite{Ulrich1, Ulrich2, Ma, Tavare, Zum, Virone, Butler,
Mit, Kaya}. The internal structure of the growing crystal, its surface morphology, and how these conditions affect kinetic attachment at the crystal face, are the explanations most often given.

Another related phenomenon is size dependent growth (SDG). This effect is primarily due to smaller crystals, say less than 2 $\mu$m in size, having higher solubility due to the Gibbs-Thomson effect and thus having slower growth rates. Another well-known size dependent effect is that of Ostwald ripening. However, ripening effects are not considered here as discussion will be restricted to the intermediate stage of crystal growth that rests in time between the induction period and any ripening phase.

Typically, in the study of the kinetics of batch crystal growth from solution one is interested in crystals of dimensions greater than 2 $\mu$m. In the past, groups have attributed certain growth rate anomalies in crystal ensembles, with sizes greater than 2 $\mu$m, to be due to SDG \cite{Jones, Nagy, Wu}. However, as pointed out by Srisanga and co-workers \cite{Ulrich2}, these effects can often be caused by the more fundamental phenomenon of GRD.

Another issue, intertwined with the kinetics and size distribution of crystals growing from solution, is the location of the greatest limitation in mass transfer during growth. Here we take the simplified view of there being one of two possible limiting restrictions. One, the case where incorporation at the crystal face is the rate limiting feature. That is, there is an overly sufficient concentration of solute near the crystal surface so that the only delay is in the incorporation of adatoms into the crystal face. Growth of this type is said to occur in the \textit{kinetc regime}. The second possibility is that the diffusion of solute in the bulk solution is much slower than incorporation at the crystal face. Crystal growth in the presence of this restriction is referred to as \textit{diffusion limited growth} (DLG). One can account for the rate of attachment at the crystal face with a surface attachment coefficient, $\kappa$, and for solute diffusion in the solution with a diffusion coefficient, $D$.

Interestingly, a DLG model can introduce a size dependent effect upon the growth rate. This comes about due to using a steady state approximation for dilute solutions that involves the crystal radius \cite{Kirkaldy}. However, as the effect does not depend upon the chemical or physical properties of the crystals themselves, we will not consider this to be an example of true SDG.

A simple continuity equation with source term has been used effectively by many groups to describe the evolution of crystals growing from supersaturated solutions or melts \cite{Barlow, Barlow2, Barlow3, Buyevich, Aoun, Ripen}. This method is often referred to as population balance. However, a more thorough approach is to use the full Fokker-Planck equation. The Fokker-Planck equation, traditionally used to describe a stochastic ensemble of some sort \cite{MCP}, has been generalized and used by many researchers to study the crystal growth process \cite{Alexandrov1, Alexandrov2, Alexandrov3}. While similar to the traditional continuity equation, a second order diffusive term is added. An expression for mass-conservation is then combined with the Fokker-Planck equation to yield a fully described integro-differential system. The rewarding aspect of solving this system is that it will yield expressions for a crystal size distribution function and the supersaturation decay. A typical solution strategy is to use a known model for the growth rate and then to assume that all crystals are born into the system as critical nuclei of zero radius, so called homogeneous nucleation. 

Often, when solving this system, the distribution function is assumed non-separable in crystal radius and time. Obviously this makes a complete closed form solution without approximations more difficult to obtain. Fortunately, several groups have shown how this can be done using suitable approximations in the resolution of certain integrals which produce supersaturation and distribution curves that are descriptive of experimental results \cite{Alexandrov1, Buyevich, Barlow}.

Other groups have been able to find exact closed form solutions for this system by assuming a separable distribution function \cite{Barlow2, Barlow3}. These results have been shown to be consistent with the experimentally observed crystal size distributions at equilibrium for insulin \cite{Schlitch}. The separable approach, although leading to a more readily obtained closed form, has the disadvantage of introducing a separation constant and yielding a distribution function that cannot have a time dependent maximum. However, at equilibrium, neither the separable nor the non-separable distribution functions are time dependent. It is the equilibrium situation that we model in this work using a separable distribution function.

In the next section, an exact closed form solution is found for our generalized Fokker-Planck equation, assuming a separable distribution function, for the cases of diffusion limited growth with and without GRD. When there is no GRD the governing equation is readily separated and integrated. The resulting equilibrium distribution function has interesting features that depend upon the values of the kinetic parameters $\kappa$ and $D$. As expected, when $D \gg \kappa$ the distribution function reduces to a form typical of growth in the kinetic regime. However, when diffusion limitation becomes significant ($D \approx \kappa$), the result predicts an equilibrium distribution with a non-zero, most probable size. We then show how the model can be used to predict the $\kappa/D$ ratio required to achieve the largest, most probable, crystal size at equilibrium. Comparisons are made with reported experimental data for insulin and protein crystal growth in aqueous environments.

For the case where GRD is assumed present, the distribution function has the intriguing result of requiring that one of the kinetic parameters in the problem be restricted to a spectrum of discrete values. Though the distribution function is found for the batch growth situation, we show how the model can be easily adapted and used to describe equilibrium size distributions report for cases of continuous crystallization. A comparison is made with size distribution data for the growth of lactose and sucrose crystals where the authors suggested that the crystal ensemble be composed of crystals with two distinct types of kinetic behavior: one slow growing the other fast \cite{Liang, Shi, Hartel}. It is shown how these two growth species can be modeled using the derived distribution function with each having a distinct value for the growth rate diffusivity coefficient.

\section{Theory}

We will consider the Fokker-Planck equation in the following form:
\begin{equation}
\frac{\partial{f}}{\partial{t}} + \frac{\partial G f}{\partial r} = \frac{\partial }{\partial r} \left(P\frac{\partial f}{\partial r}
\right)~.
\label{eq:1}
\end{equation}
Here $r$ is the radius, or characteristic dimension, of a single crystal and $t$ time. $f = f(r,t)$ gives the distribution of crystal sizes and $G$ is a linear growth rate, $G = dr/dt$. The parameter $P$, often referred to as the growth rate diffusivity, accounts for the possibility of GRD in the distribution. $P$ may be a constant or a function of $r$ and/or $t$.

Typically, a model is selected for $G$ and then the solution for Eq. (\ref{eq:1}) determined. Forms where $G$ is solely a function of time are the most common. However, in this work, we considered the case where $G = G(r, t)$. We invoke the model described by Kirkaldy and Young \cite{Kirkaldy}
\begin{equation}
G = \frac{\kappa s(t)}{1+\frac{\kappa }{D}r}~.
\label{eq:2}
\end{equation}
where $s$ is the normalized supersaturation, that is, $s(0) = 1$. We set $r = 0$ at $t = 0$. An additional mass conservation equation, mentioned in the introduction, need not be considered here as we do not seek an expression for $s(t)$.

A commonly used model for $G$ is for the case when $D \gg \kappa$ and Eq. (\ref{eq:2}) reduces to $G(t) = \kappa s(t)$. Power laws of this form are also commonly employed such as $G = ks^p$ for $p$ a positive integer. 

Such models work well for situations where the rate limiting step to growth is incorporation at the crystal face, the so-called \textit{kinetic regime}. Here we consider cases where the full range of values for the ratio $\kappa/D$ is taken into account.

\subsection{Diffusion limited growth with no GRD}

We seek a solution for $f$ in Eq. (\ref{eq:1}) for the case where $G$ is given by Eq. (\ref{eq:2}) and $P = 0$. An exact analytic, closed from solution can be found if we let $f$ be separable, that is $f = R(r)T(t)$. Using this form for $f$ in Eq. (\ref{eq:1}), along with $G$ from Eq. (\ref{eq:2}), one finds after taking derivatives and re-arranging that
\begin{equation}
\frac{1}{Ts\kappa}\frac{dT}{dt} = \frac{\kappa}{D\left(1+\frac{\kappa}{D}r\right)^2} - \frac{1}{R \left(1 + \frac{\kappa}{D}r\right)}\frac{dR}{dr}~.
\label{eq:3}
\end{equation}
Eq. (\ref{eq:3}) can be separated.

Letting the separation constant be $\lambda$ we get from the right side of Eq. (\ref{eq:3}):
\begin{equation}
\frac{\kappa}{D\left(1+\frac{\kappa}{D}r\right)^2} - \frac{1}{R \left(1 + \frac{\kappa}{D}r\right)}\frac{dR}{dr} = \lambda~.
\label{eq:4}
\end{equation}
This result can be further separated. After re-arranging Eq. (\ref{eq:4}) one is lead to
\begin{equation}
\int_0^r \frac{\kappa}{D\left(1+\frac{\kappa}{D}r \right)}dr - \lambda \int_0^r \left(1 + \frac{\kappa}{D}r \right)dr = \int_{R_o}^R \frac{dR}{R}~,
\label{eq:5}
\end{equation}
where we have let $R(0) = R_o$. Performing the integration in Eq. (\ref{eq:5}) and re-arranging one finds that
\begin{equation}
R(r) = R_o \left(1+\frac{\kappa}{D}r \right) \exp \left[-\lambda \left(r+\frac{\kappa}{2 D}r^2 \right) \right]~.
\label{eq:6}
\end{equation}

Using conservation of mass, the left side of Eq. (\ref{eq:3}) can be used to find $s(t)$ and $T(t)$ and has been described previously, \cite{Barlow2}. Here we will be interested in the system at equilibrium where $s = 0$ and $T$ is a constant. We can therefore write the distribution function at equilibrium, $f_e$, as
\begin{equation}
f_e = f_{o} \left(1+\frac{\kappa}{D}r \right) \exp \left[-\lambda \left(r+\frac{\kappa}{2 D}r^2 \right) \right]~,
\label{eq:7}
\end{equation}
where $f_{o}$ is a constant. Here $f$ gives the density of crystals per radius $r$ at time $t$. The critical nuclei is taken to be the crystal with $r = 0$, so that a homogeneous nucleation rate $J$ can be given by $J = f(0,t)h(t)$, where $h$ is some time dependent function which gives the rate of change in concentration of critical nuclei at time $t$. From Eq. (\ref{eq:7}) one sees that $f(0,0) = f_o$, so that $f_o$ gives the initial concentration of critical nuclei, or seeds, that are in the system at the end of the induction period.

\subsection{Diffusion limited growth with GRD}

In this section we solve Eq. (\ref{eq:1}) for the case where GRD is present, that is $P \ne 0$. A commonly used model for $P$ is to take it to be directly proportional to the linear growth rate \cite{Alexandrov2}. In this work we let
\begin{equation}
P = B_o \kappa s(t)~,
\label{eq:8}
\end{equation}
where $B_o$ is a constant. Assuming a separable distribution function, as discussed previously, then using Eq. (\ref{eq:8}) and Eq. (\ref{eq:2}) in Eq. ({\ref{eq:1}) one arrives at
\begin{equation}
\frac{1}{s T} \frac{dT}{dt} = \frac{B_o \kappa}{R}\frac{d^2 R}{dr^2} - \frac{\kappa}{R}
\frac{d}{dr}\frac{R}{\left(1+\frac{\kappa}{D}r \right)}~.
\label{eq:9}
\end{equation}
Again here we disregard the time dependent part of Eq. (\ref{eq:9}). Setting the separation constant to be $\tau$ and carrying out the prescribed differentiation, Eq. (\ref{eq:9}) leads to
\begin{equation}
\frac{d^2R}{dr^2} +\frac{\kappa R}{B_o D} \frac{1}{\left(1+\frac{\kappa}{D}r \right)^2} -\frac{1}{B_o \left(1+\frac{\kappa}{D}r \right)}\frac{dR}{dr} = \frac{\tau}{B_o \kappa} R~.
\label{eq:10}
\end{equation}
Using the transformation
\begin{equation}
u = 1 + \frac{\kappa}{D}r~,
\label{eq:11}
\end{equation}
Eq. (\ref{eq:10}) can be manipulated into the form
\begin{equation}
\frac{d^2R}{du^2} -\frac{D}{B_o \kappa}\frac{1}{u}\frac{dR}{du} + 
\frac{D}{B_o \kappa}\frac{1}{u^2} R - \frac{\tau D^2}{B_o \kappa^3}R = 0~.
\label{eq:12}
\end{equation}
For convenience, the following substitutions are made:
\begin{equation}
A = \frac{D}{B_o \kappa}~,
\label{eq:13}
\end{equation}
and
\begin{equation}
B = \frac{\tau A D}{\kappa^2}~.
\label{eq:14}
\end{equation}
with these substitutions Eq. (\ref{eq:10}) can be written as
\begin{equation}
\frac{d^2R}{du^2} - \frac{A}{u}\frac{dR}{du} + \left(\frac{A}{u^2} - B \right) R = 0~.
\label{eq:15}
\end{equation}
Now, according to Theorem 4.12 from Reference \cite{Bell}, the differential equation
\begin{equation}
x^2 \frac{d^2y}{dx^2} + (1-2\alpha)x \frac{dy}{dx} + \left\{\beta^2\gamma^2x^{2\gamma} + (\alpha^2 - n^2 \gamma^2)\right\} y = 0~,
\label{eq:16}
\end{equation}
has the general solution
\begin{equation}
y(x) = C_1 x^{\alpha} J_n \left(\beta x^{\gamma}\right) + C_2 x^{\alpha} Y_n \left(\beta x^{\gamma}\right)~,
\label{eq:17}
\end{equation}
where $\alpha$, $\beta$, $\gamma$, $C_1$ and $C_2$ are constants, $n$ is an integer and $J_n$ and $Y_n$ are Bessel functions of the first and second kind respectively. Eq. (\ref{eq:15}) is of the form of Eq. (\ref{eq:16}) when we make the assignments $\alpha = (1+A)/2$, $\beta = i\sqrt{B}$ and $\gamma = 1$. Using these substitutions in Eq. (\ref{eq:16}), defining
\begin{equation}
n = \frac{A-1}{2}~,
\label{eq:18}
\end{equation}
and multiplying through by $u^2$ yields
\begin{equation}
u^2 \frac{d^2 R}{du^2} + (1-2\alpha)u \frac{dR}{du} + (2\alpha - 1 - Bu^2) R = 0~.
\label{eq:19}
\end{equation}
Therefore, a general solution for Eq. (\ref{eq:15}) can be written as
\begin{equation}
R(u) = C_1 u^{\frac{1+A}{2}} J_n \left(i\sqrt{B} u\right) + C_2 u^{\frac{1+A}{2}} Y_n \left(i\sqrt{B} u\right)~.
\label{eq:20}
\end{equation}

We seek a particular solution which has the properties that $R > 0$ for all $r$ and $R \rightarrow 0$ as $r \rightarrow \infty$. These qualifications are typical of, $K_n$, the modified Bessel function of the second kind. As such, we write our particular solution in terms of $K_n$. To accomplish this, we use Theorem 4.14 from Reference \cite{Bell} which states that
\begin{equation}
K_n (x) = \frac{\pi}{2}i^{n+1}\left[i Y_n(ix) + J_n (ix) \right]~.
\label{eq:21}
\end{equation}
Making the substitutions\\\\
$x = \sqrt{B} u$~~~~and,\\\\
$C_1 = \frac{\pi}{2}i^{n+2}$~,~~~~$C_2 = \frac{\pi}{2}i^{n+1}$~,\\\\
We then arrive at the required particular solution for Eq. (\ref{eq:15}):
\begin{equation}
R(u) = R_o u^{\frac{1+A}{2}} K_n \left( \sqrt{B} u\right)~,
\label{eq:22}
\end{equation}
where $R_o$ is normalization constant. Using Eq. (\ref{eq:11}) in this gives $R(r)$:
\begin{equation}
R(r) = R_o \left(1 + \frac{\kappa}{D}r\right)^{\frac{1+A}{2}} K_n \left( \sqrt{B} \left[1 + \frac{\kappa}{D}r\right]\right)~.
\label{eq:23}
\end{equation}
As in the case of Eq. (\ref{eq:7}) we write the full distribution function at equilibrium, $f_e$, by replacing $R_o$ in Eq. (\ref{eq:23}) with $f_o$.

The requirements of Eq. (\ref{eq:18}) are discussed now. By Eq. (\ref{eq:13}) it must be that $A > 0$ so that by Eq. (\ref{eq:18}), we are restricted to $A \ge 1$ for $n$ to be an integer. Further, we must have that $A$ is an odd positive integer and $n = 0, 1, 2, 3, \cdots$.

An analytic expression for the largest most probable size at equilibrium, is not readily obtainable from Eq. (\ref{eq:23}) and thus will not be pursued here. However, the intriguing aspects of Eq. (\ref{eq:18}) will be considered and applied in modeling size distribution data reported for lactose and sucrose in the following section.

\section{Comparison with Experiment}

\subsection{DLG with no GRD}

Though the presence of GRD has been experimentally confirmed to exists in a number of systems, there seem to be cases where GRD is very nearly absent in the growth process. With only homogeneous nucleation present, a purely exponential size distribution function is predicted to form in the absence of GRD \cite{Berglund}. Such distributions have been observed for insulin crystallization \cite{Schlitch} and have been predicted for lysozyme crystallization \cite{Barlow2}. Therefore, the presence of a very nearly exponential size distribution function is indicative of a system that can be modeled as having no GRD. As observed from  Eq. (\ref{eq:6}), the model used here predicts the presence of exponential and non-exponential equilibrium distributions, for situations with no GRD, depending upon the value of the ratio $\kappa/D$.

In this sub-section, we discuss the relevant features of Eq. (\ref{eq:6}) while making some observations and comparisons with recently reported experimental data for insulin crystals grown in the kinetic regime and for protein crystals grown in diffusion limited environments.

In the kinetic regime, $\kappa/D \rightarrow 0$ so that Eq. (\ref{eq:7}) reduces to the exponential form for the equilibrium distribution of sizes; a result typical for models where the growth rate is independent of crystal size \cite{Larson}. Such distributions are also sometimes observed after batch crystallization and during continuous crystallization \cite{Schlitch, Hartel}. Experimental data from such a distribution can be used to establish a value for $\lambda$. Using the exponential distribution to compute the mean size, it can be shown that the constant $\lambda$ is related to the mean crystal size, $\langle r \rangle$, as
\begin{equation}
\lambda = \frac{1}{ \langle r \rangle}~.
\label{eq:24}
\end{equation}

When $\kappa/D$ is finite, Eq. (\ref{eq:6}) yields a non-zero first derivative. That is, unlike for the case of the kinetic regime where the most probable size in the distribution is infinitesimally small, a finite, non-zero most probable size develops in the distribution. It is instructive to visualize the distribution function, given by Eq. (\ref{eq:7}), for selected values of $\kappa/D$. Such plots are shown in Figure 1.

\begin{figure}[h]
  \centering
  \includegraphics[width=15.0 cm]{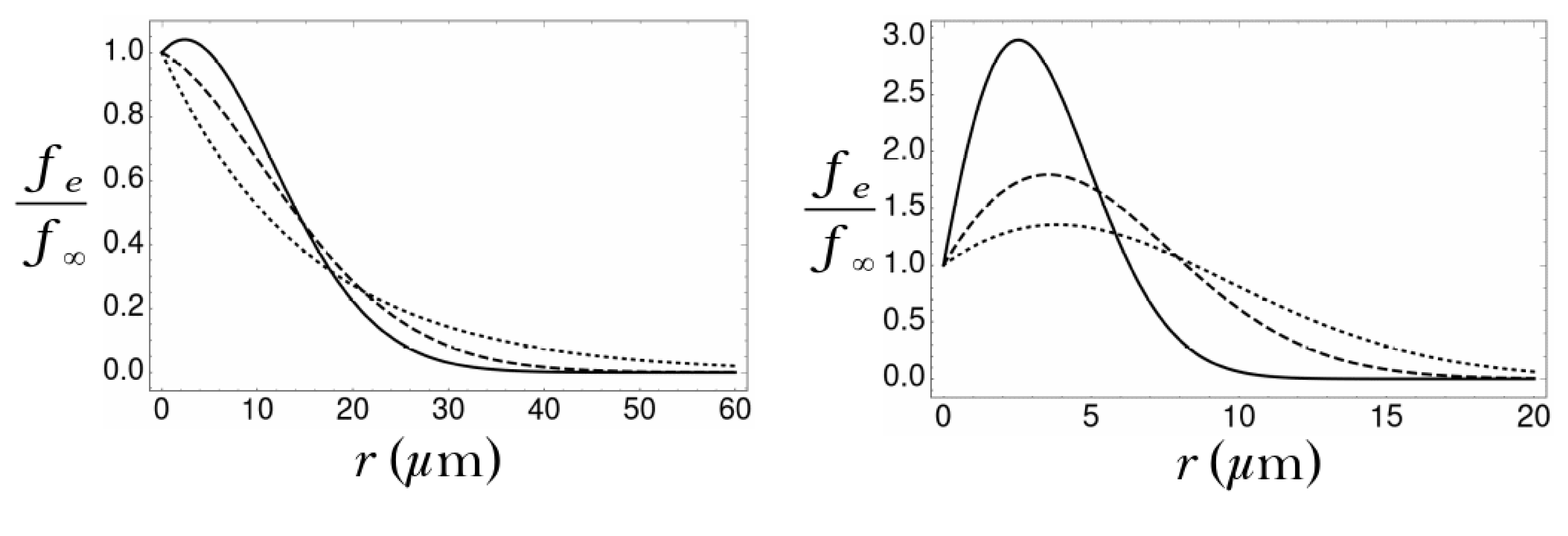}
\caption{Plots of the normalized distribution function of Eq. (\ref{eq:7}) for different values of the ratio $\kappa/D$. Figure on the left, dotted curve is for $\kappa/D$ = 0 $\mu$m$^{-1}$, dashed curve 0.05 $\mu$m$^{-1}$ and solid curve 0.1 $\mu$m$^{-1}$. For the curves in the Figure on the right, dotted curve has 0.25 $\mu$m$^{-1}$, dashed curve 0.5 $\mu$m$^{-1}$ and solid curve 1.5 $\mu$m$^{-1}$. $\lambda$ is set to 0.065 $\mu$m$^{-1}$ taken from Reference \cite{Schlitch}, a value typical for protein crystals batch grown from solution in the kinetic regime.} 
\end{figure}

From the curves in Figure 1 it can be seen that, for fixed $\lambda$, the most probable size in the distribution varies with $\kappa/D$. For $\kappa/D = 0$ the most probable size is not defined but, as $\kappa/D$ increases this values becomes finite and finally reaches the largest possible most likely size. With further increase in $\kappa/D$ the most probable size decreases. 

It is useful to have an expression for the most probable size, $r_{mp}$, as a function of $\kappa/D$ at fixed $\lambda$. Taking the first derivative of Eq. (\ref{eq:6}), setting this to zero and solving for $r$ one arrives at an expression for $r_{mp}$:
\begin{equation}
r_{mp} = \frac{1}{\kappa/D}\left(\sqrt{\frac{\kappa}{D\lambda}}-1 \right)~.
\label{eq:25}
\end{equation}
A plot of Eq. (\ref{eq:25}) is given in Figure 2 for one of the cases considered in Figure 1.

\begin{figure}[t]
  \centering
  \includegraphics[width=12.0 cm]{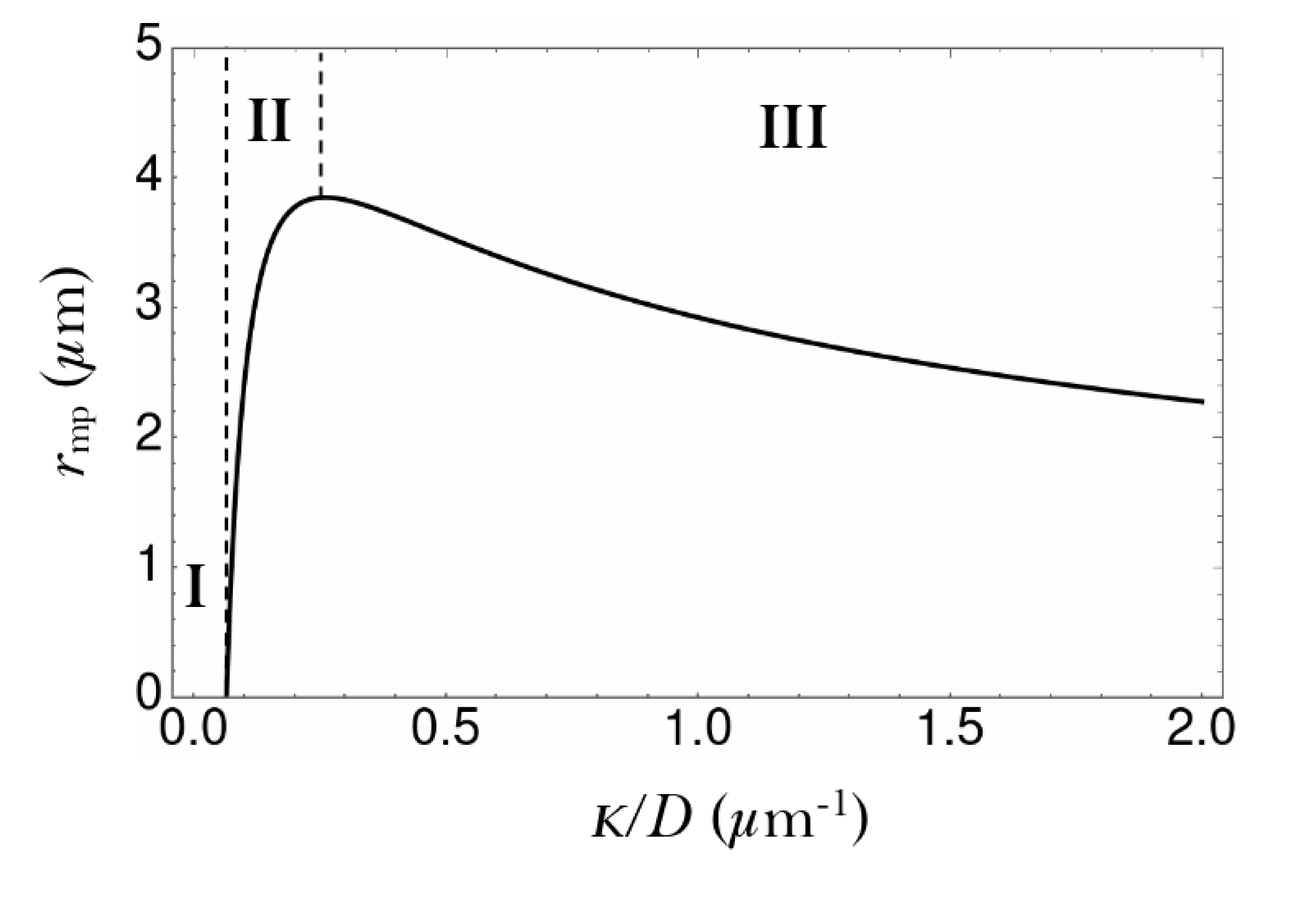}
\caption{Plot of the most probable crystal size at equilibrium vs. the ratio $\kappa/D$ for $\lambda$ = 0.065 $\mu$m$^{-1}$. Region I is the kinetic zone, region II an intermediate zone while region III is the dendritic zone.  } 
\end{figure}

Figure 2 reveals two important points in the $r_{mp}$ versus $\kappa/D$ curve. Firstly
\begin{equation}
\frac{\kappa}{D} = \lambda~~~\text{for}~~r_{mp} = 0~,
\label{eq:26}
\end{equation}
and secondly
\begin{equation}
\frac{\kappa}{D} = 4\lambda~~~\text{for}~~r_{mp}~~\text{a maximum}~.
\label{eq:27}
\end{equation}
This result, along with Eq. (\ref{eq:24}), predicts that the diffusion limited distribution with the largest most probable size will occur when $\kappa/D$ equals 4 on the average crystal radius at equilibrium after growth under kinetic regime conditions. For example, using the data from the caption to Figure 2 for the insulin studied in Reference \cite{Schlitch}, we find this model predicts that the distribution with the largest, most probable size occurs when $\kappa/D = 0.26$ $\mu$m$^{-1}$. Using a reported value for the diffusion coefficient of insulin in water at 32 $^{\text{o}}$ C of $D = 2.0 \times 10^{-10}$ m$^2$/s \cite{Evangelos} in this ratio for $\kappa/D$ one finds the corresponding attachment coefficient to be $5.2 \times 10^{-5}$ m/s or about 52 $\mu$m/s. Now, should an attachment coefficient of this magnitude be challenging to obtain, one might then reduce the solute diffusion coefficient for obtaining the distribution with the largest most probable size. For example, several groups have reported on the use of solution gel additives to manipulate the equilibrium size distribution of protein crystals \cite{Mihoko, Ruiz}.

It should be noted that $r_{mp}$ at the point where $\kappa/D = 4/\langle r \rangle$ is not nearly the largest crystal to be expected in the equilibrium distribution. However, this model predicts that the conditions which yield this value for $\kappa/D$ then lead to an equilibrium distribution where the majority of the crystals are very nearly of one particular size, an outcome often desirable in industry.

The equilibrium morphology of crystals, grown under diffusion limited conditions, is a topic of continuing research. Reports in the literature seem to indicate that proteins crystallized in the kinetic regime form cubic, dodecahedral and rhombohedral structures with maximum sizes of around 70-300 $\mu$m \cite{Schlitch, Ootaki, Ferr}. Such crystals can well be described geometrically by one characteristic distance and a constant form factor. These growth conditions correspond to region I in Figure 2, a region we entitle the kinetic zone.

There are reports for protein crystals grown under highly diffusion limited conditions, for example, when $D \approx 10^{-12}$ m$^2$/s. Such cases have been reported for green fluorescent protein \cite{Kono} and model human IgG grown in vivo and cellulo \cite{Hasegawa1, Hasegawa2}. Here rod shaped crystals grow to around 50-70 $\mu$m in length and about $10 ~\mu$m in diameter. This growth situation likely corresponds to region III in Figure 2, which we name as the dendritic zone. If the ratio of the length of the rod to its diameter is assumed constant, then these crystals can be geometrically described, within this model, by their diameter and a constant form factor. Therefore, in region III, $r_{mp}$ would correspond to the rod diameter.

Less is known by the authors concerning the morphology of protein crystals formed in region II of Figure 2, labeled as the intermediate zone. This intermediate region could correspond to cases where, as an example, $D$ for insulin has a value on the order of $5.0 \times 10^{-11}$ m$^2$/s as is reported for insulin in pancreatic tissue \cite{Buchwald}.

\subsection{DLG with GRD}

In this sub-section, the results above for DLG with GRD are used to describe some experimental results where GRD was thought to have occurred during the growth process. Cases of continuous crystallization of lactose and sucrose will be considered. The results leading to Eq. (\ref{eq:23}), though derived for the batch case, are easily adapted for the case of continuous crystallization.

The continuous crystallization process considered here is continuous mixed suspension, mixed product removal (CMSMPR) \cite{Liang, Wu}. In this growth process the supersaturation is maintained at a constant level and crystals grow in the solution for a certain residence time $t_r$. Crystals when removed, have the same distribution of sizes as those in the crystallizer. In this way, so called \textit{iso-kinetic} conditions are maintained, a sort of quasi-equilibrium. 

The theoretical results derived above can be adapted for such cases by setting $s(t) = s_o$ where $s_o$ is a constant. With this change in Eq. (\ref{eq:8}) the right side of Eq. (\ref{eq:9}) is unchanged. Therefore $R(r)$ in this case is still given by Eq. (\ref{eq:23}). The time dependent side of Eq. (\ref{eq:9}) yields
\begin{equation}
T = T_o e^{s_o \tau t_r}~,
\label{eq:28}
\end{equation}
where, $T_o$, $s_o$ and $t_r$ are all constants.

Size distribution data has been reported for the continuous crystallization of lactose \cite{Liang, Shi} and sucrose \cite{Hartel} where the resulting data cannot be fully described by an exponential curve as there was found to be a surprisingly large number of small crystals in the distribution. Liang and co-workers have suggested that these results be described with a two component model where there is a distinct group of small, slow growing crystals and a smaller group of larger, fast growing crystals in the system \cite{Liang}. This hypothesis led us to use the results discussed previously to describe this phenomenon in terms of a system containing two distinct crystal groups each with a distinct growth rate diffusion coefficient $P_n$. 

To compare Eq. (\ref{eq:23}) to reported experimental data, values must be established for $D$, $\kappa$, $B_o$ and $\tau$. The constant $\tau$ can be further clarified by considering the solution to Eq. (\ref{eq:9}) when there is no DLG, that is $\kappa/D \rightarrow 0$, and yet $B_o \ne 0$. In this case the right side of Eq. (\ref{eq:9}), and the separation constant $\tau$, lead to a second order linear differential equation with constant coefficients. The solution that suits our conditions is
\begin{equation}
R(r) = R_o e^{\alpha r}~,
\label{eq:29}
\end{equation}
where
\begin{equation}
\alpha = \frac{1}{2B_o} \left(1-\sqrt{1+\frac{4B_o \tau}{\kappa} } \right)~.
\label{eq:30}
\end{equation}
In the limit of $B_o \rightarrow 0$, $\alpha$ reduces to $-\tau/\kappa$. Relating this to the mean size in the distribution after growth in the kinetic regime, as done in deriving Eq. (\ref{eq:24}), leads to
\begin{equation}
\tau = \frac{\kappa}{\langle r \rangle}~.
\label{eq:31}
\end{equation}

Now set $P_n = {B_o}_n \kappa$, where ${B_o}_n$ now has a spectrum of discrete values according to the index $n$ as does $P_n$, the growth rate diffusivity of order $n$. Now, using Eq. (\ref{eq:13}) in Eq. (\ref{eq:18}) we find that $P_n$ can be related to the solute diffusion coefficient $D$ as
\begin{equation}
P_n = \frac{D}{2n +1}~.
\label{eq:32}
\end{equation}
Using Eqs. (\ref{eq:13}), (\ref{eq:18}) and (\ref{eq:32}), in Eq. (\ref{eq:23}), the constant ${B_o}_n$ is eliminated in $R(r)$. This leaves $R$ in terms of the most convenient unknown parameters:
\begin{equation}
R(r) = R_o  \left(1 + \frac{\kappa}{D}r\right)^{{n+1}} K_n \left[ \frac{\sqrt{\tau(2n+1)D}}{\kappa} \left(
1 + \frac{\kappa}{D}r\right) \right]~.
\label{eq:33}
\end{equation}

To compare Eq. (\ref{eq:33}) to experimental data an estimate must first be made for the constant $\tau$. By Eq. (\ref{eq:31}) it must be that $\sqrt{\tau}/\kappa = 1/\sqrt{\kappa \langle r \rangle}$. This relationship is defined for the case where there is no GRD. As will be shown, the model described above predicts significant GRD for the faster growing ensembles of larger crystals while less GRD for slow growing systems of smaller crystals. A typical mean size for the reported lactose distributions studied here is around 10 $\mu$m. Therefore, since crystals are assumed to start from nuclei of sub-micron size, we speculate that for a growth run with no GRD, involving the materials studied here, the mean size in such an ensemble might be in the neighborhood of 1 $\mu$m and as will be shown, $\kappa$, is on the order of 1 $\mu$m/s. We therefore make the assignment that $1/\sqrt{\kappa \langle r \rangle} \approx 1$ $\sqrt{\text{s}}/\mu$m. In the case considered here for sucrose, we make the same assignment as an approximately doubling in average final size as compared to lactose, is compensated by a halving of $\kappa$.

In the papers reporting crystal size distribution during continuous crystallization of lactose \cite{Liang}, alpha-lactose monohydrate \cite{Shi} and sucrose \cite{Hartel}, the data is given in a plot as $\ln f$ versus $r$. In all three cases the data is along the straight line of $\ln f$ versus $r$ for large sizes but $\ln f$ values rise well above this line for small sizes. Assuming that the distribution is composed of two distinct crystal types, we take the data for the faster growing system to be the data that lies along the line of $\ln f$ versus $r$. The smaller, slower growing ensemble begins at the size where data begins to fall off this straight line. This transition location is obvious in the figures from References \cite{Liang, Shi, Hartel}. We assign the faster growing ensemble to be the $n = 0$ mode in Eq. (\ref{eq:33}) and adjust $\kappa$ to achieve a fit to the reported data. One of the experimental data points, in approximately the center of the set in each case, is used to establish a value for the constant $R_o$. This result for lactose crystals is shown in Figure 3a for the lactose crystal growth case given in Reference \cite{Liang}.

Leaving the parameters $\kappa$, $D$ and $1/\sqrt{\kappa \langle r \rangle}$ at the same value, the index $n$ is increased, (in this case to 5), a new value for $R_o$ is set using an experimental data point and this version of Eq. (\ref{eq:33}) is compared to data from Reference \cite{Liang} from the slow growing section of the distribution. This results is shown in Figure 3b.

Now, the sum of the curve for the distribution of slow growing crystals plus the distribution for the fast growing crystals, as used in Figures 3a and 3b, should yield a reasonable fit to the complete data set from the distribution. This is indeed the case and the resulting curve is shown fit to the data in Figure 3c. In Figure 3d the log of Eq. (\ref{eq:33}) versus $r$, using parameter values as set for Figures 3a, 3b and 3c, is shown for selected values of $n$ with each line pinned to the same data point near the middle of the plot. Also shown in Figure 3d is the log of the experimental data. It can be seen how the values of $n$ define a family of nearly straight lines.

This process is then repeated for the reported data for alpha-lactose monohydrate and sucrose with the results given in Figures 4 and 5 respectively.  Data used to generate the curves are given in the captions.

\begin{figure}[h]
  \centering
  \includegraphics[width=14.0 cm]{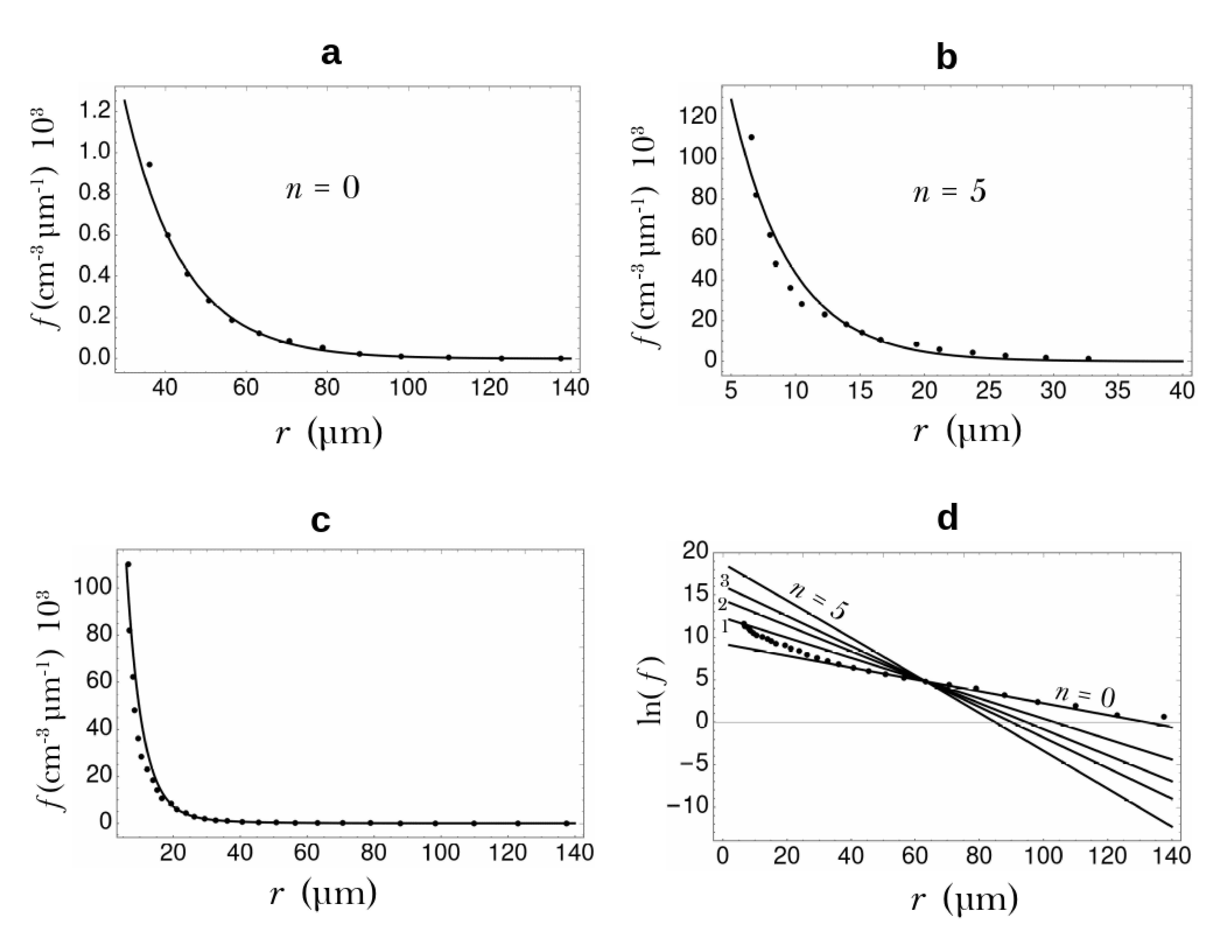}
\caption{Plots of Eq. (\ref{eq:33}) and size distribution data for lactose from Reference \cite{Liang}. Figure \textbf{a} shows Eq. (\ref{eq:33}) for the case where $n = 0$ fit to data for the fast growing crystals. $\kappa = 1.48 ~\mu$m/s and $D = 430.0 ~\mu$m$^2$/s \cite{Hober}. Figure \textbf{b} shows Eq. (\ref{eq:33}) for the case where $n = 5$ fit to data for the slow growing crystals. Figure \textbf{c} shows the sum of the $n = 0$ and $n = 5$ curves being used as a fit to all of the distribution data. Figure \textbf{d} shows plot of the log of Eq. (\ref{eq:33}) for $n = 0, 1, 2, 3, 5$ compared to the log of the experimental data.} 
\end{figure}

\begin{figure}[h]
  \centering
  \includegraphics[width=14.0 cm]{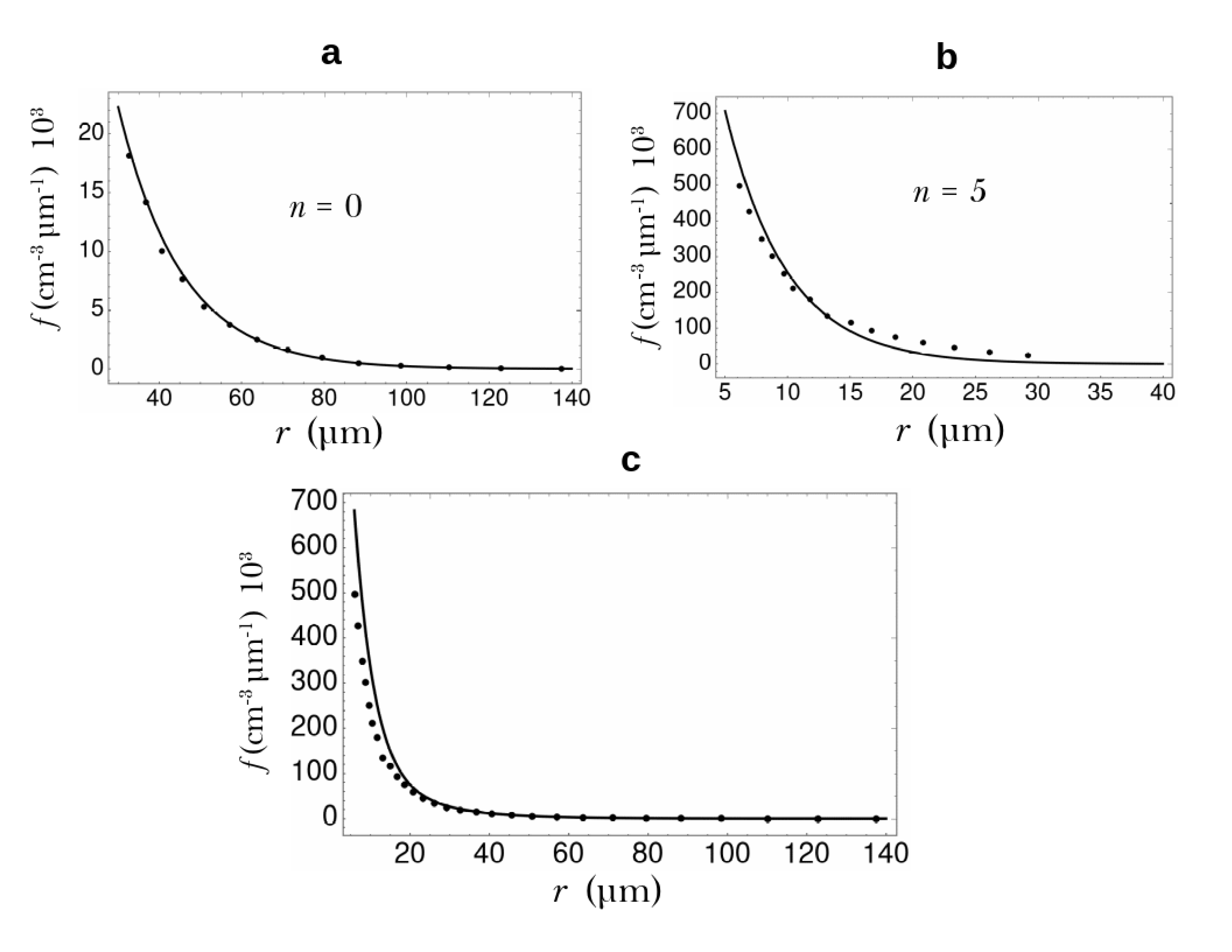}
\caption{Plots of Eq. (\ref{eq:33}) and size distribution data for lactose from Reference \cite{Shi}. Figure \textbf{a} shows Eq. (\ref{eq:33}) for the case where $n = 0$ fit to data for the fast growing crystals. $\kappa = 1.38 ~\mu$m/s and $D = 430.0 ~\mu$m$^2$/s \cite{Hober}. Figure \textbf{b} shows Eq. (\ref{eq:33}) for the case where $n = 5$ fit to data for the slow growing crystals. Figure \textbf{c} shows the sum of the $n = 0$ and $n = 5$ curves being used as a fit to all of the distribution data.} 
\end{figure}

\begin{figure}[h]
  \centering
  \includegraphics[width=14.0 cm]{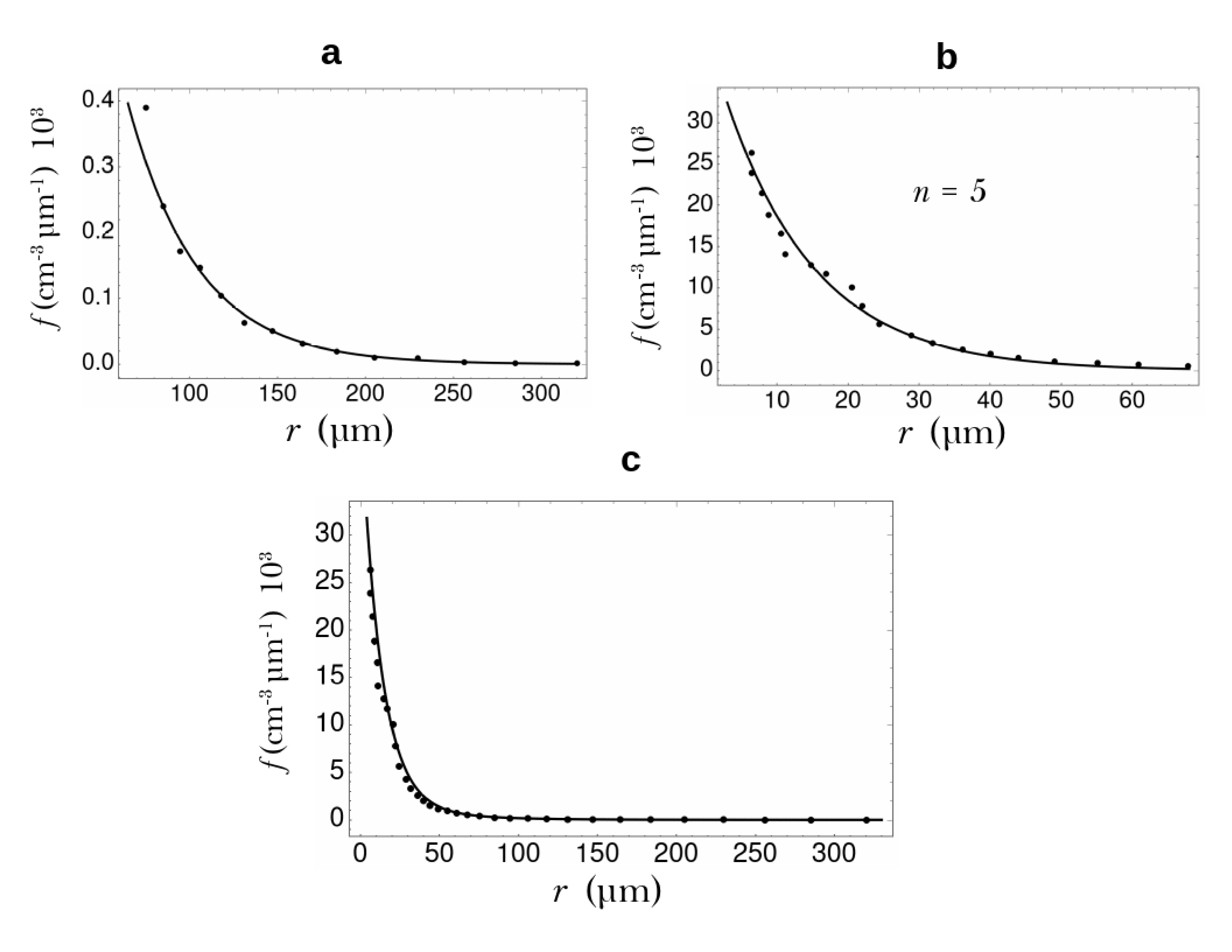}
\caption{Plots of Eq. (\ref{eq:33}) and size distribution data for sucrose from Reference \cite{Hartel}. Figure \textbf{a} shows Eq. (\ref{eq:33}) for the case where $n = 0$ fit to data for the fast growing crystals. $\kappa = 0.55 ~\mu$m/s and $D = 458.6 ~\mu$m$^2$/s \cite{Atkins}. Figure \textbf{b} shows Eq. (\ref{eq:33}) for the case where $n = 5$ fit to data for the slow growing crystals. Figure \textbf{c} shows the sum of the $n = 0$ and $n = 5$ curves being used as a fit to all of the distribution data.} 
\end{figure}

\clearpage
\section{Conclusion}

In this report, a model for crystal DLG with and without GRD was given and compared with experimental data. In the case with no GRD, the model yields a simple, convenient, closed form analytic result for the equilibrium distribution of sizes. This distribution function can then be used to derive an expression for the most probable size in the equilibrium distribution as a function of the ratio $\kappa/D$. A plot of the most probable size versus $\kappa/D$ suggests the labeling of three distinct growth regimes. The first is the the kinetic zone where there is no non-zero maximum in the distribution and $\kappa/D$ is small, then as $\kappa/D$ increases, a non-zero maximum develops in the distribution. Eventually, this maximum reaches the largest, most probable size. The values for $\kappa/D$ beyond the largest, most probable size we label as the dendritic zone and speculate that rod-like, or other elongated shaped crystals will form in these solution conditions due to strong diffusion limiting conditions. Less is known about the intermediate zone between the kinetic and dendritic zones but we speculate that crystal growth in this region could be typical of conditions in living tissue fluids.

Finally, a second order Fickian term is added to the model so that GRD can be considered. On solving this system, one arrives at a differential equation whose solution can be written in terms of the modified Bessel function of the second kind and another factor involving the independent variable. Interestingly, this result requires that one of the parameters in the system have only discrete values. We let the growth rate diffusivity be given in this way and show how the result can be used to describe size distribution data, reported for the crystal growth of lactose and sucrose, by assuming that the system is composed of two components, each with a distinct value for the growth rate diffusivity coefficient.

It should be noted that in the report by Liang and co-workers \cite{Liang} growth rates were reported in the neighborhood of 10$^{-2}$ $\mu$m/s. The values for $\kappa$ used here for fitting purposes, and listed in the captions to Figures 3, 4 and 5, are ten to one hundred times larger. However, we propose that it is not necessary for values of $\kappa$, in this application, to be on the order of a crystal growth rate for one of the components. As both slow and fast growing modes were modeled, within the framework of this theory, with identical values for $\kappa$, it then must act as a sort of effective attachment coefficient, $\kappa_{eff}$, for the system of distribution functions containing all orders of $n$. However, this effective rate is not built up from a sum of equivalent contributions throughout the orders as the individual growth rates decrease with increasing $n$. That is,
\begin{equation}
\kappa_{eff} = \sum_{n=0}^{\infty}\kappa_n~,
\label{eq:34}
\end{equation}
where $\kappa_n \rightarrow 0$ as $n \rightarrow \infty$.

The physical mechanism that might underlie a system showing the presence of a spectrum of kinetic modes, is a mystery to the authors and requires further research. Before speculating on this issue, we discuss some of the other relevant results of this model. In the experimental reports considered here \cite{Liang, Shi, Hartel}, the authors suggest that the growth process occurred in the kinetic regime of growth. However, even when $\kappa/D$ is less than one, say around $10^{-3}$ $\mu$m$^{-1}$, a value indicative of the kinetic regime, $\kappa/D$ is perhaps comparatively small but not zero. As seen from work in a previous section, this makes all of the difference since when $\kappa/D = 0$ the notion of multiple, distinct, growth rate diffusivity values for the system, disappears. This indicates, (see Eq. (\ref{eq:32})), that for cases where $\kappa/D$ is finite, which in reality would be all growth cases, the bulk solute diffusivity is intimately connected with the growth rate diffusivity; as if the limitation of the solute molecule to transit the solution is mirrored in GRD. Or, from another perspective, the fluctuations involved in solute diffusion are similar in nature to those of GRD. 

As for the discrete nature of a kinetic parameter as predicted by the model presented here, physical phenomena displaying the characteristics of partitioning are not unheard of in this field. The faceted nature of growing crystals is an excellent example of this and it is well known that growing crystals can have face specific growth rates \cite{Ma, Herden, Herden2}. In particular, Dincer and co-workers \cite{Dincer1} have reported experimental measurements that seem to show that different faces of growing lactose crystals exhibit growth rates that are integer multiples of one another. Mitrovi\'{c} and co-workers \cite{Mit} have shown that for KDP crystals grown from aqueous solution that three distinct crystal groups appear, each group having a distinct mean growth rate. 

\section{CRediT authorship contribution statement}

\textbf{D. Barlow:} Conceptualization, Investigation, Formal Analysis, Writing-review and editing. \textbf{K. Monaghan:} Formal Analysis, Writing-review and editing.

\section{Declaration of competing interest}
The authors have no conflicts of interest to report.

\section{Data availability}
All data used to arrive at conclusions in this report are listed in the report.

\section{Acknowledgements}

The authors wish to thank the College of Arts and Sciences at The University of the South for funding support.


\end{document}